\documentstyle[12pt,epsfig]{article}

\def\be{\begin{equation}}
\def\ee{\end{equation}}
\def\bea{\begin{eqnarray}}
\def\eea{\end{eqnarray}}
\def\ba{\begin{array}}
\def\ea{\end{array}}
\def\bi{\begin{itemize}}
\def\ei{\end{itemize}}
\def\bc{\begin{center}}
\def\ec{\end{center}}
\def\bfr{\begin{flushright}}
\def\efr{\end{flushright}}
\def\noi{\noindent}
\def\non{\nonumber}
\def\ol{\overline}
\def\wt{\widetilde}
\def\tanb{\tan\beta}
\def\cotb{\cot\beta}
\def\gsim{\lower.7ex\hbox {$\;\stackrel{\textstyle>}{\sim}\;$}}
\def\lsim{\lower.7ex\hbox {$\;\stackrel{\textstyle<}{\sim}\;$}}
\def\ZZ{\hbox {\it Z\hskip -4.pt Z}}

\newcommand{\raltitle}
{Phenomenology of a New Minimal Supersymmetric\\
Extension of the Standard Model}

\newcommand{\ralauthor}
{A. Dedes$^1$, C. Hugonie$^1$, S. Moretti$^1$ and K. Tamvakis$^2$} 

\newcommand{\raladdress}
{$^1$Rutherford Appleton Laboratory, \\
Chilton, Didcot, Oxon, OX11 0QX, UK. \medskip \\
$^2$Physics Department, Theory Division, \\ 
University of Ioannina, GR451 10, Greece.} 

\newcommand{\ralabstract}
{We study the phenomenology of a new Minimally-extended Supersymmetric
Standard Model ({\em nMSSM}) where a gauge singlet superfield is added
to the MSSM spectrum. The superpotential of this model contains no dimensionful
parameters, thus solving the $\mu$-problem of the MSSM. A global discrete
$R$-symmetry, forbidding the cubic singlet self-interaction, imposed on the
complete theory, guarantees its stability with respect to generated
higher-order tadpoles of the singlet and solves both the domain wall and
Peccei-Quinn axion problems. We give the free parameters of the model and
display some general constraints on them. A particular attention is devoted to
the neutralino sector where a (quasi-pure) singlino appears to be {\it always}
the LSP of the model, leading to additional cascades, involving the NLSP $\to$
LSP transition, compared with the MSSM. We then present the upper bounds on the
masses of the lightest and next-to-lightest -- when the lightest is an
invisible singlet -- CP-even Higgs bosons, including the full one-loop and
dominant two-loop corrections. These bounds are found to be much higher than
the equivalent ones in the MSSM. Finally, we discuss some phenomenological
implications for the Higgs sector of the nMSSM in Higgs production at future
hadron colliders.}

\newcommand{\ralpreprint}
{RAL-TR-2000-036}

\begin{document}

\begin{titlepage}

\bfr\ralpreprint\efr
\vspace{3cm}

\bc
 {\large \bf \raltitle} \\
 \vspace{2cm}
 {\bf \ralauthor} \\
 \bigskip
 {\it \raladdress} \\
 \vspace{2cm}
 {\bf Abstract} \\
 \bigskip
\ec
\ralabstract
\setcounter{page}{0}

\end{titlepage}

\section{Introduction}

Supersymmetry provides a well defined framework for the study of physics beyond
the Standard Model (SM). Its main motivation has been the special properties of
supersymmetric (susy) theories with respect to the hierarchy problem. In
addition, the low energy data support unification of the gauge couplings in the
susy case, in contrast to what happens in the SM scenario. Another interesting
feature of susy models is that the breaking of the electroweak (EW) symmetry
can be radiatively triggered by the largeness of the top quark mass \cite{RSB}.
The Minimal Supersymmetric Standard Model (MSSM) \cite{NHK} is defined by
promoting each standard field into a superfield, doubling the Higgs fields and
imposing $R$-parity conservation. Due to the non-observation of superpartners
of the standard particles, supersymmetry has to be broken at a scale $M_{susy}$
not larger than $O$(TeV), so that it still provides a natural solution to the
hierarchy problem.  Unfortunately, a phenomenologically acceptable realization
of EW symmetry breaking in the MSSM requires the presence of the so-called
$\mu$-term, a direct susy mass term for the Higgs fields, with values of the
(theoretically arbitrary) parameter $\mu$ close to $M_{susy}$ or $M_{W}$, when
its natural value would be either $0$ or $M_P$. Of course, there exist
explanations for an $O(M_W)$ value of the $\mu$-term, alas, all in extended
settings \cite{mupb}.

The more or less straightforward solution to the $\mu$-problem is to promote
the $\mu$-parameter into a field whose vacuum expectation value (v.e.v.) is
determined, like the other scalar field v.e.v.'s, from the minimization of the
scalar potential along the new direction \cite{NMSSM1,NMSSM2,NMSSM3,NMSSM4}.
Naturally, it is expected to fall in the range of the other v.e.v.'s, i.e., of
order $O(M_{susy})$. Such a superfield has to be a singlet under the SM gauge
group. In order to avoid introducing new scales into the model one should stick
to dimensionless couplings at the renormalizable level. This can be achieved by
imposing a $\ZZ_3$ symmetry on the renormalizable part of the superpotential.
The resulting model, the Next-to-Minimal Supersymmetric Standard Model (NMSSM),
has the following superpotential:

\be
W = \lambda S H_1 H_2 + \frac{\kappa}{3} S^3 + ... \label{supot}
\ee

\noi where the dots stand for the usual quark and lepton Yukawa couplings (cf
eq.~(\ref{superpot})). The $\ZZ_3$ symmetry is spontaneously broken at the EW
scale when the Higgs fields get a non-zero v.e.v. It is well known, however,
that the spontaneous breaking of such a discrete symmetry results in disastrous
{\it{cosmological domain walls}}, unless this symmetry is explicitly broken by
the non-renormalizable sector of the theory. Domain walls can be tolerated if
there is a discrete-symmetry-violating contribution to the scalar potential
larger than the scale $O$(1 MeV) set by nucleosynthesis \cite{walls}. Heavy
fields interacting with the standard light fields generate in the effective
low-energy theory an infinite set of non-renormalizable operators of the light
fields scaled by powers of the characteristic mass-scale of the heavy sector
($M_P$, $M_{GUT}$,...). These terms appear either as $D$-terms in the
K\"{a}hler potential or as $F$-terms in the superpotential. It is known however
that gauge singlet superfields {\it do not obey decoupling}
\cite{mupb,singlets}, so that, when supersymmetry is either spontaneously or
softly broken, in addition to the suppressed non-renormalizable terms, they can
in general give rise to a {large} tadpole term in the potential proportional to
the heavy scale: $M_{susy}^2 M_P (S+S^*)$. Technically, the tadpole is
generated through higher-order loop diagrams in which the non-renormalizable
interactions participate as vertices together with the renormalizable ones. A
discrete global symmetry like the one discussed above would forbid this term
but would lead to the appearance of disastrous domain walls upon its
unavoidable spontaneous breakdown. The generated large tadpole reintroduces the
hierarchy problem, since due to its presence the singlet v.e.v. gets a value
$\langle S \rangle^2 \sim M_{susy} M_P$. It appears that $N=1$ supergravity,
spontaneously broken by a set of hidden sector fields, is the natural setting
to study the generation of the destabilising tadpoles. A thorough analysis
carried out in Ref.~\cite{Abel} shows that the only harmful non-renormalizable
interactions are either {\it even superpotential} terms or {\it odd K\"{a}hler
potential} ones. In addition, operators with more than six powers of the
cut-off in the denominator are harmless. Finally, a tadpole diagram is
divergent only if it contains an {\it odd} number of `dangerous' vertices.

The solution of the $\mu$-problem in the framework of the NMSSM could be
rendered a viable one if the destabilization problem were circumvented. What is
needed is a suitable symmetry that forbids the dangerous non-renormalizable
terms and allows only for tadpoles of order $M_{susy}^3 (S+S^*)$. This symmetry
should at the same time allow for a large enough $\ZZ_3$-breaking term in the
scalar potential in order to destroy the unwanted domain walls
\cite{Tamvakis1}.

An alternative approach is to impose a symmetry which, although it does not
forbid the dangerous non-renormalizable terms, only allows for higher-order
tadpole graphs that give a $n$-loop-suppressed term $\frac{1}{(16\pi^2)^n}
M_{susy}^2 M_P (S+S^*)$. A case of particular interest is when the cubic
self-interaction for the singlet in eq.~(\ref{supot}) is forbidden by the
symmetry. Actually, it should be noted that if the underlying theory is a Grand
Unified one (GUT), although a candidate for the singlet exists, a cubic term
does not arise\footnote {In $E_6$ for example, matter and Higgs fields are
contained in the {\bf 27} representation together with a singlet. Although the
standard trilinear singlet-Higgses term is present in the {\bf 27}$^3$
coupling, no singlet cubic term arises. The same is true for $E_6$ embedings of
$SO(10)$ and $SU(5)$.}. On the other hand, this case is truly minimal in the
sense that, apart from promoting the $\mu$-parameter into a field, no new
renormalizable terms appear in the superpotential. Of course, a substitute is
needed for the twofold role played by the cubic term, namely, its contribution
to the mechanism generating the v.e.v. of $S$ through the soft susy breaking
terms and the breaking of the {\it Peccei-Quinn symmetry} present when $\kappa
= 0$. This role can be played by the tadpole. (Note that this is not included
in the $\kappa \rightarrow 0$ limit of the existing NMSSM analyses, which up to
now have ignored the tadpole term \cite{NMSSM2,NMSSM3,cascades1}.) Recently, a
viable solution along these lines was proposed based on {\it discrete
$R$-symmetries} \cite{Tamvakis2}. The renormalizable superpotential for this
{\it new minimal supersymmetric extension} of the Standard Model (nMSSM) is
given by

\be
W = \lambda SH_1H_2 + Y_u Q U^c H_1 + Y_d Q D^c H_2 + Y_e L E^c H_2.
\label{superpot}
\ee

\noi Apart from the usual Baryon and Lepton number, it possesses two additional
global continuous symmetries, namely, an anomalous Peccei-Quinn symmetry
$U(1)_{PQ}$ with charges

\be
Q (-1) ,\, U^c (0) ,\, D^c (0) ,\, L (-1) ,\, E^c (0) ,\, H_1 (1) ,\, H_2 (1)
,\, S (-2)
\ee

\noi and a non-anomalous $R$-symmetry $U(1)_R$ with charges

\be
Q (1) ,\, U^c (1) ,\, D^c (1) ,\, L (1) ,\, E^c (1) ,\, H_1 (0) ,\, H_2 (0) ,\,
S (2) .
\ee

\noi One of the solutions worked out consists in imposing the discrete
sub-symmetry $\ZZ_{5 R}$ of the $U(1)_{R'}$ combination $R'=3R+PQ$ on the
complete theory, including non-renormalizable operators. The charges under
$\ZZ_{5 R}$ are

\bea
(H_1,H_2) & \rightarrow & \alpha (H_1,H_2), \non \\
(Q,L) & \rightarrow & \alpha^2 (Q,L), \non \\
(U^c,D^c,E^c) & \rightarrow & \alpha^3 (U^c,D^c,E^c), \\
S & \rightarrow & \alpha^4 S, \non \\
\cal{W} & \rightarrow & \alpha \cal{W}, \non
\eea

\noi where $\alpha = e^{2i\pi/5}$. An adequately suppressed linear term is
generated at six-loop level by combining the non-renormalizable K\"{a}hler
potential terms $\lambda_1 S^2 H_1 H_2 / M_P^2 + h.c.$ and $\lambda_2 S
(H_1H_2)^3 / M_P^5 + h.c.$ with the renormalizable superpotential term $\lambda
S H_1 H_2$:

\be
V_{tadpole} \sim \frac{1}{(16\pi^2)^6} \lambda_1\lambda_2\lambda^4 M_{susy}^2
M_P (S+S^*). \label{tadpole}
\ee

\noi This tadpole has the desired order of magnitude $O(M_{susy})$ if
$\lambda_1\lambda_2\lambda^4 \sim 10^{-3}$.

The goal of this paper is the phenomenological study of this nMSSM, where a
gauge singlet superfield is added to the MSSM spectrum and a global $\ZZ_{5 R}$
symmetry is imposed on the complete theory, resulting in the superpotential of
eq.~(\ref{superpot}) and the tadpole term of eq.~(\ref{tadpole}). In section 2
we review the general properties of the parameter space of the model.
Phenomenological aspects of the nMSSM are addressed in section 3. The
neutralino sector, including the (quasi-pure) singlino, is studied in some
detail. Also bounds on CP-even Higgs masses versus their couplings to gauge
bosons are displayed along with Higgs production cross sections at future
hadron colliders. Section 4 contains our main conclusions.

\section{Model set-up}

The tree-level Higgs scalar potential, namely, the potential which contains
the scalar fields $H_1=(H_1^0,H_1^-)$, $H_2=(H_2^+,H_2^0)$ and
$S$, has the form :

\bea
V^{(0)} & = & V_F + V_D + V_{soft} + V_{tadpole} , \label{V0} \\
V_F & = & |\lambda|^2 \biggl [ \biggl (|H_1|^2+|H_2|^2 \biggr ) |S|^2 + |H_1|^2
|H_2|^2 \biggr ] \non \\
& - & |\lambda|^2 \biggl ( H_1^{0*}H_2^{0*}H_1^-H_2^+ + h.c \biggr ) \\
V_D & = & \frac{g_1^2+g_2^2}{8} \biggl [ |H_1|^2 - |H_2|^2 \biggr ]^2 +
\frac{g_2^2}{2}|H_1^\dagger H_2|^2 , \\
V_{soft} & = & m_{H_1}^2 |H_1|^2 + m_{H_2}^2 |H_2|^2 + m_S^2 |S|^2 \non \\
& + & \biggl ( \lambda A_{\lambda} S H_1 H_2 + h.c \biggr ) .
\eea

\noi In what follows, we shall assume a phenomenological point of view and
write the generated tadpole as

\be
V_{tadpole} \equiv \xi^3 (S+S^*) ,
\ee

\noi where $\xi$ is treated as a free parameter.

In order to obtain the correct upper limits on the Higgs boson masses (cf.
section \ref{secH}) radiative corrections to the tree-level potential have to
be considered. Let us introduce a scale $Q \sim M_{susy}$ and assume that
quantum corrections involving momenta $p^2 \gsim Q^2$ have been evaluated, e.g.
by the integration of the Renormalization Group Equations (RGEs) of the
parameters from initial values at the GUT scale down to the scale $Q$. One is
then left with the computation of quantum corrections involving momenta $p^2
\lsim Q^2$. The effective potential $V_{ef\!f}$ can be developed in powers of
$\hbar$ or loops as

\be
V_{ef\!f} = V^{(0)} + V^{(1)} + V^{(2)} + \ldots .
\ee

\noi The tree-level potential $V^{(0)}$ is given by eq.~(\ref{V0}). The
one-loop corrections to the effective potential read as

\be
V^{(1)} = \frac{1}{64\pi^2} \, \mbox{STr} M^4 \left [ \ln \left ( 
\frac{M^2}{Q^2} \right ) - \frac{3}{2} \right ] ,
\ee

\noi where $M^2$ is the field dependent squared mass matrix (in our analysis,
we take only top/stop loops into account). Next, we consider the dominant
two-loop corrections. These will be numerically important only for large susy
breaking terms compared to the Higgs v.e.v.'s $h_i$, hence we can expand in
powers of $h_i$. Since the terms quadratic in $h_i$ can be absorbed into the
tree-level soft terms, we just consider the quartic terms, and here only those
which are proportional to large couplings: terms $\sim \alpha_s h_t^4$ and
$\sim h_t^6$. Finally, taking only leading logarithms (LLs) into account, the
expression for $V^{(2)}$ reads

\be
V^{(2)}_{LL} = 3 \left ( \frac{h_t^2}{16\pi^2} \right )^2 h_2^4 \left (
32\pi\alpha_s - \frac{3}{2} \, h_t^2 \right ) t^2 ,
\ee

\noi where $t \equiv \ln \left (\frac{Q^2}{m_t^2} \right )$, $m_t$ being the
top quark mass. One-loop corrections to the tree-level relations between bare
parameters and physical observables, once reinserted in the one-loop effective
potential, also appear as two-loop effects. These are: corrections to the
kinetic terms of the Higgs bosons, which lead to a wave function
renormalization factor $Z_{H_2}$ in front of the $D_\mu H_2 D^\mu H_2$ term
with, to order $h_t^2$

\be
Z_{H_2} = 1 + 3 \frac{h_t^2}{16\pi^2} \, t ; \label{zh}
\ee

\noi and corrections to the top quark Yukawa coupling with, to orders $h_t^2$,
$\alpha_s$

\be
h_t(m_t) = h_t(Q) \left ( 1 + \frac{1}{32\pi^2} \left ( 32\pi\alpha_s -
\frac{9}{2} \, h_t^2 \right ) t \right ) .
\ee

In general, the parameters $\lambda$, $A_\lambda$ and $\xi$ could be complex.
However, by redefining the fields $H_2$ (or $H_1$) and $S$, one can always get
-- without loss of generality -- that $\lambda A_\lambda, \xi^3 \in R$. Note
that in the NMSSM with the cubic singlet superpotential term $\frac{1}{3}
\kappa S^3$ one has to further {\it assume } that the combination $\lambda
\kappa^*$ (or, equivalently, $A_\lambda/A_\kappa$) is real
\cite{NMSSM2,NMSSM3}. By $SU(2)_L\times U(1)_Y$ gauge invariance one can get
rid of the phase of $H_1$, by taking $\langle H_1^- \rangle = 0$ and $h_1
\equiv \langle H_1^0 \rangle \in R^+$. One can then show that the condition for
a local minimum with $\langle H_2^+ \rangle = 0$ is equivalent to a positive
mass squared for the charged Higgs. It has been proven that a sufficient
condition is $\lambda < g_2$ \cite{Pokorski} which, as we shall see below, is
always verified in the universal case. By taking $h_2 \equiv \langle H_2^0
\rangle = \rho_2 e^{i\phi_2}$, $s \equiv \langle S \rangle = \rho_0
e^{i\phi_0}$ and minimizing the complete (two-loop) effective potential with
respect to $\phi_0$ and $\phi_2$, we find that there is one and only one global
vacuum for which the two phases relax to zero, i.e., $\phi_0 = \phi_2 = 0$.
This implies that there is {\em no spontaneous CP-violation}. Therefore one can
choose $h_1 \in R^+$ and $h_2 , s \in R$. This result distinguishes the nMSSM
from the usual NMSSM where loop corrections can generate spontaneous
CP-violation \cite{CP}.

The soft terms of the model can be constrained by requiring universality at
the GUT scale. The independent parameters of the model are then a universal
gaugino mass $M_{1/2}$ (always positive in our convention), a universal mass
for the scalars $m_0^2$, a universal trilinear coupling $A_0$ (either positive
or negative), the (positive) Yukawa coupling $\lambda_0$ at the scale
$M_{GUT}$ and the tadpole coefficient $\xi$. The (well-known) value of the
$Z$-boson mass fixes one of these parameters with respect to the others, so
that we end up with {\it four free parameters at the GUT scale}, i.e., as many
as in the MSSM with universal soft terms. In principle, one could choose the
same set of free parameters as in the MSSM, i.e., $M_{1/2}$, $m_0^2$, $A_0$
and $\tanb (\equiv \frac{h_2}{h_1})$, with $\lambda$, $s$ and $\xi$ being
determined by the three minimization equations, demanding also radiative
electroweak symmetry breaking~\cite{RSB}. However, this appears to be a
non-trivial issue, as $\lambda$ also influences the running of the RGEs of the
soft parameters between the GUT and the EW scale. In other terms, one would
need a lot of fine tuning of the dimensionful $A_0$ in order to get a
dimensionless parameter, $\lambda$, of the desired value at the EW scale.
Therefore, in the case of universality, we conveniently adopt in our numerical
analysis the following input parameters: $m_0^2/M_{1/2}$, $A_0/M_{1/2}$,
$\xi/M_{1/2}$ and $\lambda_0$ ($\tanb$ and $s$ being calculated from the
minimization of the potential and the overall scale $M_{1/2}$ fixed by
$M_Z$).

If one requires the absence of a Landau singularity for $\lambda$ below the GUT
scale, one obtains an upper bound on $\lambda$ at the EW scale. This upper
bound depends on the value of the top quark Yukawa coupling $h_t$, i.e., on
$\tanb$ (cf. fig.~1). Requiring furthermore, universality at the GUT scale, one
ends up with the more restrictive constraint $\lambda \lsim .3$, higher values
leading to unphysical global minima of the effective potential.

\begin{figure}[ht]
\bc
\epsfig{file=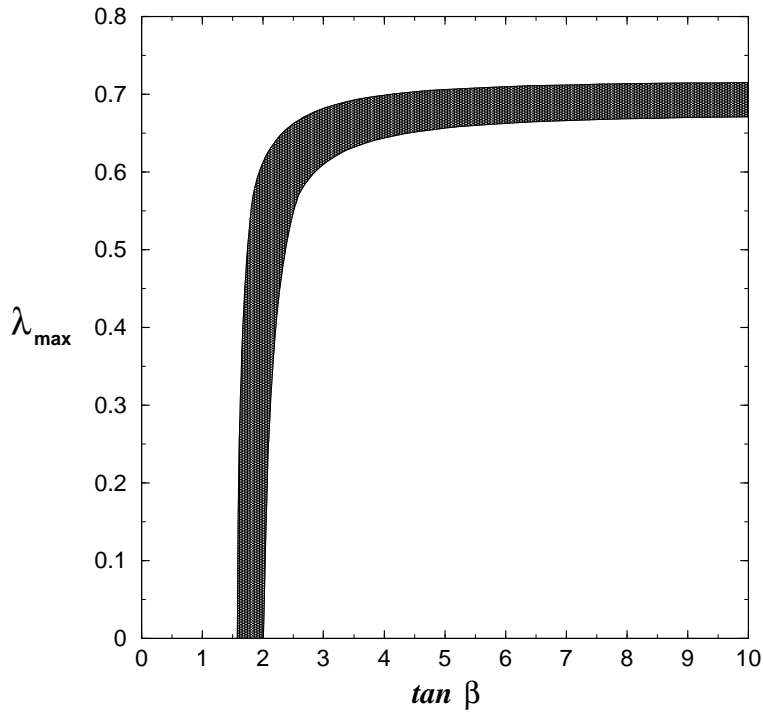,width=10cm}
\ec
\caption{Upper bound on $\lambda$ as a function of $\tanb$ for $m_t^{pole} =
173.8 \pm 5.2$ GeV \cite{PDG}. The width of the curve is due to the
uncertainty on $m_t^{pole}$.}
\end{figure}

Let us now briefly address the problem of Charge and Color Breaking (CCB)
minima. The most dangerous CCB direction involves the trilinear coupling $h_e
A_e E_{R,1} L_1 H_1$ where $h_e$ denotes the electron Yukawa coupling ($\sim
10^{-5}$), $E_{R,1}$ is the right-handed selectron, $L_1$ the left-handed
slepton doublet of the first generation, and $H_1$ the corresponding Higgs
doublet. From the absence of a non-trivial minimum of the scalar potential in
the $D$-flat direction $|E_{R,1}| = |L_1| = |H_2|$, the following inequality
among the soft susy breaking terms can be derived \cite{NMSSM1}:

\be
A_e^2 < 3 \left ( m_E^2 + m_L^2 + m_1^2 \right ) , \label{CCB}
\ee

\noi where $m_E^2$, $m_L^2$ and $m_1^2$ are the soft susy breaking mass terms
associated with the three fields above. If the inequality (\ref{CCB}) is
violated, the fields develop v.e.v.'s of $O \left ( A_e/h_e \right )$ and the
depth of the minimum is of $O \left ( A_e^4/h_e^2 \right )$. Accordingly,
(\ref{CCB}) has to be imposed at a scale $Q \sim A_e/h_e \sim 10^7$~GeV.
Assuming universal soft terms at the GUT scale, (\ref{CCB}) then becomes
\cite{NMSSM3}:

\be
\left ( A_0 - 0.5 M_{1/2} \right )^2 < 9 m_0^2 + 2.67 M_{1/2}^2 .
\ee

\noi So-called UFB directions (for Unbounded From Below, which actually never
occurs in the universal case) can also be considered. Assuming universality for
the soft terms, the absence of a global minimum in these directions typically
implies \cite{UFB1,UFB2}

\be
\frac{m_0}{M_{1/2}} \gsim 1 \label{UFBbound} .
\ee

\noi However, the tunnelling rate from the standard EW minimum to a UFB one is
in general quite small \cite{UFB1}, so that this constraint can be avoided if
one is ready to assume that the standard EW vacuum is metastable.

\section{Phenomenological aspects of the nMSSM}

\subsection{Singlino LSP and additional cascades}

The nMSSM contains additional gauge singlet states in the Higgs sector (one
neutral CP-even and one CP-odd state) and in the neutralino sector (a two
component Weyl fermion). These states are mixed with the corresponding ones of
the MSSM, and the physical states have to be obtained from the diagonalization
of the mass matrices in each sector. In the basis $\left ( \wt{B}, \wt{W}^3,
\wt{H}_1^0, \wt{H}_2^0, \wt{S} \right )$ the (symmetric) neutralino mass matrix
reads as

\be
{\cal M}^0 = \left(
\ba{ccccc}
M_1 & 0 & -g_1h_1/\sqrt{2} & g_1h_2/\sqrt{2} & 0 \\
& M_2 & g_2h_1/\sqrt{2} & -g_2h_2/\sqrt{2} & 0 \\
& & 0 & \lambda s & \lambda h_2 \\ &
& & 0 & \lambda h_1 \\
& & & & 0
\ea
\right) . \label{neumm}
\ee

\noi Note that the diagonal singlino ($\wt{S}$) mass term is zero. Furthermore,
the singlino mixings (with Higgsinos) are proportional to $\lambda$, which, as
remarked earlier, turns out to be quite small, especially in the universality
scenario. Consequently, the singlino state appears to be an almost pure singlet
state with a very small mass, so that it is always the lightest supersymmetric
particle (LSP) of the model. Actually, in the universal case, we find: few MeV$
\lsim m_{\wt{S}} \lsim 3$ GeV, with a singlet component $\gsim 99\%$. This
state has only small couplings to the gauge bosons and to the other sparticles.
(We have explicitly checked that its contribution to the invisible $Z$-boson
width is $< 4.2 $ MeV \cite{PDG}.) Thus, the production cross sections of the
singlino are small and it seems to be nearly impossible to observe this
particle in any experiment, this rendering the nMSSM apparently similar to
the ordinary MSSM. However, as the singlino is the LSP of the model, it will
appear at the end of all sparticle decay chains, giving rise to additional
cascades compared with the MSSM signals. Such additional cascades are common to
many supersymmetric extensions involving singlets \cite{cascades1,cascades2}.
It should be noticed, however, that unlike in the NMSSM, where the singlino LSP
scenario requires strong constraints on the parameter space (i.e., $M_{1/2} \gg
m_0, A_0$) \cite{cascades1}, the singlino is {\it always} the LSP in the
nMSSM.

Besides, by assuming universality at the GUT scale, we find that the
next-to-lightest supersymmetric particle (NLSP) is always the second lightest
neutralino, which turns out to be a quasi-pure bino ($\wt{B}$). Depending on
the region of the parameter space under scrutiny, the following channels can
play a role in the NLSP $\to$ LSP cascade:

\bi
\item $\wt{B} \to \wt{S}\nu\ol{\nu}$ (sneutrino/$Z$ exchange) giving an
invisible cascade;

\item $\wt{B} \to \wt{S}l^+l^-$ (slepton/$Z$ exchange) where the leptons could
be mainly $\tau$'s, the stau being lighter than the other sleptons;

\item $\wt{B} \to \wt{S}q\ol{q}$ (squark/$Z$ exchange) the branching ratio
being quite small ($\lsim 10\%$), as the squarks are usually heavy;

\item $\wt{B} \to \wt{S}Z$ if the $\wt{B}$ is heavy enough;

\item $\wt{B} \to \wt{S}S$ where $S$ is a light quasi-pure singlet Higgs
boson, decaying to $b\ol{b}$ or $\tau\ol{\tau}$ depending on its mass;

\item $\wt{B} \to \wt{S}\gamma$ through loops.
\ei

\noi The properties of these cascades have been analyzed in detail in
Ref.~\cite{cascades1} for the case of the NMSSM and most of the results can
equally apply to the case of the nMSSM. As for experimental searches, high
multiplicity events have been under investigation already in the context of
models with gauge mediated supersymmetry breaking \cite{GMSB} or with
$R$-parity violation \cite{Rp}. In principle, small $\lambda$'s ($\lsim
10^{-4}$) could give rise to a delayed NLSP $\to$ LSP transition, i.e. a
displaced neutral vertex \cite{cascades1,cascades2}. However, such values of
$\lambda$ are disfavoured if one wants the tadpole term of eq.~(\ref{tadpole})
to be large enough, so that displaced neutral vertices are not expected as
typical signatures of the nMSSM.

\subsection{Higgs couplings and mass bounds} \label{secH}

The Higgs sector of the nMSSM consists of three CP-even neutral states,
denoted by $S_i$ with masses $m_1<m_2<m_3$, plus two CP-odd neutral states,
labelled as $P_i$ with masses $m'_1<m'_2$. The tree-level mass matrix for the
CP-even states in the basis $(Re H_1^0, Re H_2^0, Re S)$ reads

\be
{\cal M}_S^2 =
\left( \ba{ccc}
g^2 h_1^2 - \lambda s A_\lambda \tanb & (2 \lambda^2-g^2) h_1 h_2 + \lambda s
A_\lambda & \lambda (2 \lambda s h_1 + A_\lambda h_2) \\
& g^2 h_2^2 - \lambda s A_\lambda \cotb & \lambda (2 \lambda s h_2 + A_\lambda
h_1) \\
& & -\lambda^2 A_\lambda \frac{h_1 h_2}{\lambda s} - \lambda
\frac{\xi^3}{\lambda s} \ea \right) \label{cpeven}
\ee

\noindent where $g^2 = (g_1^2+g_2^2)/2$. In the reminder of this section, we
study the upper bounds on the lightest CP-even states with general soft susy
breaking terms, in a non-universal scenario. By taking into account the full
one-loop and the dominant two-loop top/stop corrections displayed in section
2, and assuming $h_i \ll M_{susy}$, one obtains the following upper limit on
the lightest CP-even Higgs mass:

\bea
m_1^2 & \leq & M_Z^2 \left ( \cos^2 \! 2\beta + \frac{2\lambda^2}{g_1^2+g_2^2}
\, \sin^2 \! 2\beta \right ) \left ( 1 - \frac{3h_t^2}{8\pi^2} \, t \right )
\non \\
& & + \frac{3h_t^2(m_t)}{4\pi^2} \, m_t^2(m_t) \sin^2 \! \beta \left (
\frac{1}{2} \, \wt{X}_t + t + \frac{1}{16\pi^2} \left ( \frac{3}{2} \, h_t^2 -
32\pi\alpha_s \right ) ( \wt{X}_t + t ) t \right ) \label{bound1}
\eea
where
\bea
\wt{X}_t & \equiv & 2 \, \frac{\wt{A}_t^2}{M_{susy}^2} \left ( 1 -
\frac{\wt{A}_t^2}{12M_{susy}^2} \right ), \label{Xt} \\
\wt{A}_t & \equiv & A_t - \lambda s \cotb ,
\eea
$A_t$ being the top trilinear soft term.

\begin{figure}[ht]
\bc
\epsfig{file=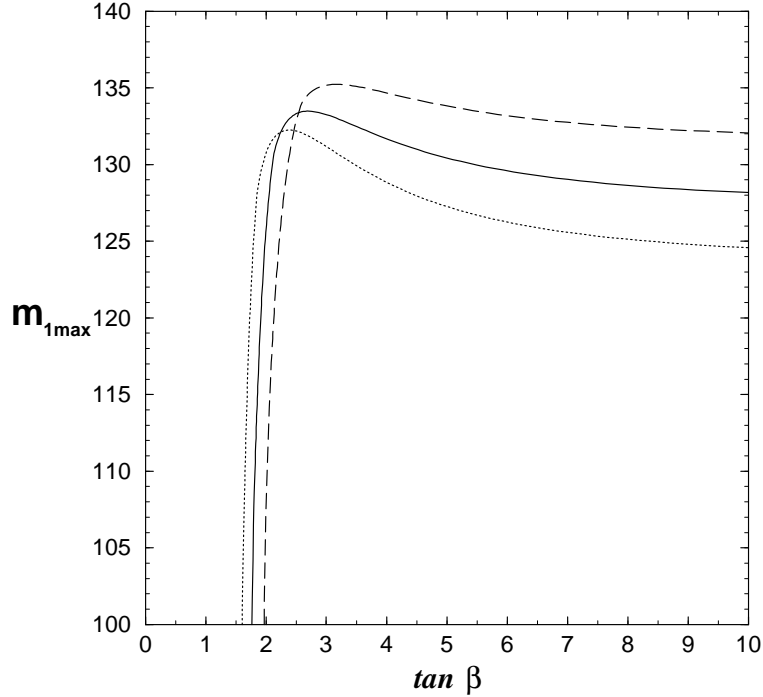,width=10cm}
\ec
\caption{Upper bound on $m_1$ [GeV] versus $\tanb$ for $m_t^{pole} = 173.8 \pm
5.2$ GeV (straight, dashed, dotted line respectively), and $M_{susy} \leq 1$
TeV.} \end{figure}

The only difference between the MSSM bound \cite{2loop} and eq.~(\ref{bound1})
is the `tree-level' contribution $\sim \lambda^2 \sin^2 \! 2\beta$. This term
is important for moderate values of $\tanb$. Hence, the maximum of the
lightest Higgs mass in the nMSSM is not obtained for large $\tanb$  values,
like in the MSSM, rather for moderate ones (cf. fig.~2). In contrast, the
radiative corrections are identical in the nMSSM and in the MSSM. In
particular, the linear dependence in $\wt{X}_t$ is the same in both models.
Hence, from eq.~(\ref{Xt}), the upper bound on $m_1^2$ is maximised for
$\wt{X}_t=6$ (corresponding to $\wt{A}_t=\sqrt{6} M_{susy}$, the 'maximal
mixing' case) and minimized for $\wt{X}_t=0$ (corresponding to $\wt{A}_t=0$,
the 'no mixing' case).

However, the upper limit on $m_1$ is not necessarily physically relevant,
since the coupling of the lightest CP-even Higgs boson to the $Z$-boson can be
very small. Actually, this phenomenon can also appear in the MSSM, if
$\sin^2(\beta-\alpha)$ is small. In this case though, the CP-odd
Higgs boson $A$ is necessarily light ($m_A \sim m_h < M_Z$ at tree-level) and
the process $e^+e^-\to Z \rightarrow h A$ can be used to cover this region of
the MSSM parameter space. In the nMSSM, a small gauge boson coupling of the
lightest Higgs $S_1$ is usually related to a large singlet component, in which
case no (strongly coupled) light CP-odd Higgs boson is available. Hence, Higgs
searches in the nMSSM have to possibly rely on the search for the second
lightest Higgs scalar $S_2$.

Let us define the reduced coupling $R_i$ as the square of the $ZZS_i$ coupling
divided by the corresponding standard model Higgs - $Z$ boson coupling

\be
R_i = (S_{i1}\cos\!\beta + S_{i2}\sin\!\beta)^2 ,
\ee

\noi where $S_{i1}, S_{i2}$ are the $H_1, H_2$ components of the CP-even Higgs
boson $S_i$, respectively. Evidently, we have $0 \leq R_i \leq 1$ and
unitarity implies

\be 
\sum_{i=1}^3 R_i = 1 . \label{Rsum} 
\ee 

We are interested in upper bounds on the two lightest CP-even Higgs bosons
$S_{1,2}$. These are obtained in the limit where the third Higgs, $S_3$, is
heavy and decoupled, i.e., $R_3 \sim 0$ (this scenario is similar the
so-called decoupling limit in the MSSM: the upper bound on the lightest Higgs
$h$ is saturated when the second Higgs $H$ is heavy and decouples from the
gauge bosons). In this limit, we have $R_1 + R_2 \simeq 1$.

In the regime $R_1 \geq 1/2$, experiments will evidently first discover the
lightest Higgs (with $m_1 \leq 133.5$ GeV for $m_t^{pole} = 173.8$ GeV and
$M_{susy} = 1$ TeV). The `worst case scenario' in this regime corresponds to
$m_1 \simeq 133.5$ GeV and $R_1 \simeq 1/2$: the presence of a Higgs boson
with these properties has to be excluded in order to test this part of the
parameter space of the nMSSM.

In the regime $R_1 < 1/2$ (i.e. $1/2 < R_2 \leq 1$) the lightest Higgs may
escape detection because of its small coupling to gauge bosons, and it may be
easier to look for the second lightest Higgs $S_2$. In fig.~3 we show the
upper limit on $m_2$ as a function of $R_2$ as a thin straight line. For $R_2
\rightarrow 1$ (i.e. $R_1 \rightarrow 0$), the upper limit on $m_2$ is
actually given by the previous upper limit on $m_1$, even if the corresponding
Higgs boson is the second lightest one. For $R_2 \rightarrow 1/2$, on the
other hand, $m_2$ can be as large as 190 GeV. However, one finds that the
upper limit on $m_2$ is saturated when the mass $m_1$ of the lightest Higgs
boson tends to 0. Clearly, one has to take into account the constraints from
Higgs boson searches which apply to reduced couplings $R < 1/2$, i.e., lower
limits on $m_1$ as a function of $R_1 \simeq 1 - R_2$, in order to obtain
realistic upper limits on $m_2$ versus $R_2$. The dotted curves in fig.~3 show
the upper limit on $m_2$ as a function of $R_2$ for different fixed values of
$m_1$ (as indicated on each curve). They can be used to obtain upper limits on
the mass $m_2$, in the regime $R_1 < 1/2$, for arbitrary experimental lower
limits on the mass $m_1$ versus $R_1$. For each value of the coupling $R_1$,
which would correspond to a vertical line in fig.~3, one has to find the point
where this vertical line crosses the dotted curve associated to the
corresponding experimental lower limit on $m_1$. Joining these points by a
curve leads to the upper limit on $m_2$ as a function of $R_2$. We have
indicated  as a thick straight line in fig.~3 the present LEP-II bounds
\cite{LEP2}, which give, in the `worst case' scenario, an upper limit on $m_2$
of $\simeq$ 160 GeV for $R_2 \simeq1/2$.

\begin{figure}[ht]
\bc
\epsfig{file=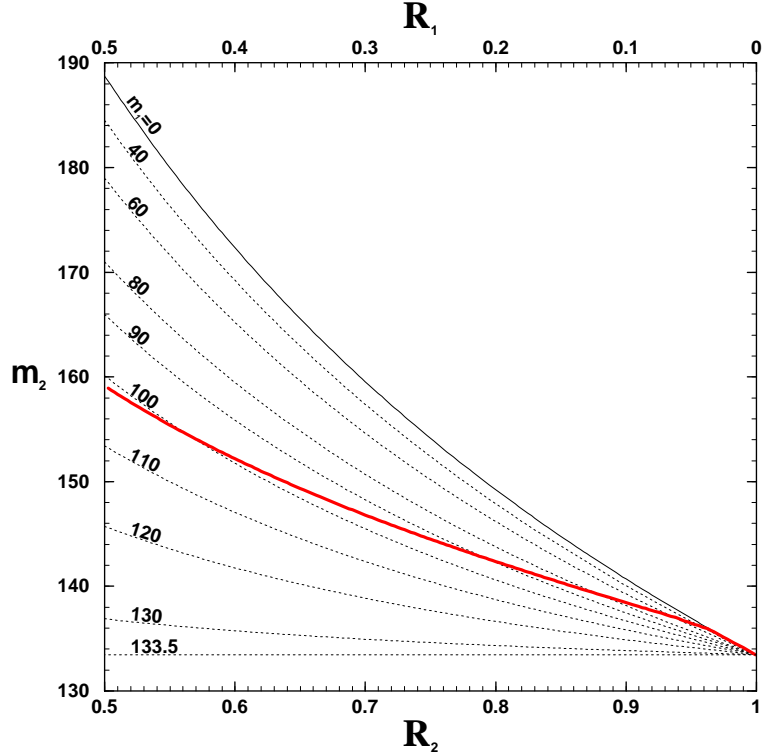,width=10cm}
\ec
\caption{Upper limits on the mass $m_2$ (in GeV) against $R_2$, for different
values of $m_1$ (as indicated on each line in GeV), assuming $m_t^{pole} =
173.8$ GeV and $M_{susy} \leq 1$ TeV. $R_1=1-R_2$ is shown on the the top
axis. The thick straight line corresponds to LEP-II lower limits on $m_1$ vs.
$R_1$.}
\end{figure}

Lower experimental limits on a Higgs boson with $R > 1/2$ restrict the allowed
regime for $m_2$ (for $R_2 > 1/2$) in fig.~3 from below. The present lower
limits on $m_2$ from LEP-II are not visible in fig.~3, since we have only
shown the range $m_2 > 130$ GeV. Possibly Higgs searches at Tevatron Run II
will push the lower limits on $m_2$ upwards into this range. This would be
necessary if one aims at an exclusion of this regime of the nMSSM. Then,
lower limits on the mass $m_2$ -- for any value of $R_2$ between $1/2$ and 1
-- of at least 133.5 GeV are required. The precise experimental lower limits
on $m_2$ as a function of $R_2$, which would be needed to this end, will
depend on the achieved lower limits on $m_1$ as a function of $R_1$ in the
regime $R_1 < 1/2$.

In principle, from eq.~(\ref{Rsum}), one could have $R_2 > R_1$ with $R_2$ as
small as $1/3$. However, in the regime $1/3 < R_2 < 1/2$, the upper bound on
$m_2$ as a function of $R_2$ for different fixed values of $m_1$ can only be
saturated if $R_1 = R_2$. It is then sufficient to look for the lightest Higgs
$S_1$ (i.e. for a Higgs boson with a coupling $1/3 < R < 1/2$ and a mass $m
\lsim 133.5$ GeV) to cover this region of the parameter space of the nMSSM.

One can notice that these results are the same as those obtained in the NMSSM
\cite{last}. This comes from the fact that the non-singlet part of the CP-even
mass matrix (\ref{cpeven}) as well as the singlet/non-singlet mixing terms are
the same in both models, giving the same upper limit on $m_1$,
eq.~(\ref{bound1}) and fig.~2. Even though they are not similar, the CP-even
singlet mass term (the $3\times3$ element in eq.~(\ref{cpeven})) is `free' in
both models, i.e., it can take any value between 0 and $\sim$ 1 TeV. This
explains why the curves displayed in fig.~3 are the same in both models, the
upper bounds on $m_2$ steaming from degenerate singlet/non-singlet mass terms,
degeneration lifted by the mixings (off-diagonal $1\times3$ and $2\times3$
terms in eq.~(\ref{bound1})).

Thus, the phenomenological potential of the nMSSM is not less exciting in
the Higgs sector than it is in the sparticle sector. On the one hand, the
necessary (but not sufficient) condition for testing the complete parameter
space of the nMSSM is to rule out a CP-even Higgs boson with a coupling $1/3
< R < 1$ and a mass below 135~GeV. On the other hand, the sufficient condition
(i.e., the precise upper bound on $m_2$ versus $R_2$) depends on the achieved
lower bound on the mass of a `weakly' coupled Higgs (with $0 < R < 1/2$) and
can be obtained from fig.~3. At the Tevatron Run II this would probably
require an integrated luminosity of up to 30 fb$^{-1}$ \cite{r15}. If this
cannot be achieved, one has to wait for the advent of the LHC, in order to
know whether the nMSSM is actually realized in nature.

Imposing universality of the soft terms at the GUT scale, CCB constraint
(eq.~(\ref {CCB})) and LEPII experimental constraints on Higgs masses
\cite{LEP2} one finds the more restrictive bound $m_1 < 122$ GeV under the
same hypotheses ($m_t^{pole} = 173.8 \pm 5.2$ GeV and $M_{susy} < 1$ TeV) and
$m_2 < 135$ GeV in the case where $S_1$ is mainly singlet ($R_1 < R_2$).

An intriguing example of possible evidence of the nMSSM at future
hadron-hadron colliders is the following. Recall that a crucial signature for
an intermediate Higgs boson at both the Tevatron and the LHC is the one
produced via $q\bar q'\to W^\pm~{\mathrm{Higgs}}$ (with a smaller contribution
from $q\bar q\to Z~{\mathrm{Higgs}}$ as well), with the gauge vector yielding
high transverse-momentum and isolated leptons and the Higgs scalar decaying
into $b\bar b$ pairs. Now, let us imagine that Higgs searches in this mode
have finally revealed the evidence of a light scalar Higgs resonance, but no
further Higgs states are detected up to well above the EW scale. (In this
respect, the LHC is a better example to illustrate, as compared to the
Tevatron, because of its much extended scope in mass.) This scenario could be
realized in the MSSM, if the latter is in the above mentioned decoupling
regime, where one has $m_{H^\pm}\approx m_H\approx m_A\gg M_Z$ and the
strength of the lightest Higgs boson couplings to the gauge vector bosons
$W^\pm$ and $Z$ approaches unity, $R_h\equiv\sin^2(\beta-\alpha)\simeq1$.
Indeed, this phenomenology can also be realized in the nMSSM when, e.g.,
$R_1=1$ and $R_2=0$\footnote{Further notice that, no matter their actual mass
value, pseudoscalar neutral Higgs states of either model cannot be produced
via the above two processes, as their coupling to $W^\pm$ and $Z$ vectors is
prohibited at tree-level.}. Under these circumstances, the mass of the $h$
scalar would only depend on $\tanb$ and the minimum of the Higgs
production cross section is obtained, in both susy models, in correspondence
of the maximum Higgs mass. Fig.~4 shows this dependence for both the MSSM and
the nMSSM at the LHC, with $\sqrt s=14$ TeV. (The trend is very similar at
the Tevatron, where $\sqrt s=2$ TeV.) Over a broad range in $\tanb$, the
production rates of the latter are significantly below those of the former,
with a minimum at $\tanb \simeq 2.7$ (for $m_t=173.8$ GeV), owning to the
peculiar $\tanb$ dependence of the maximum value of $m_1$, as seen in
fig.~2 (recall instead that the maximum mass of the lightest Higgs boson of
the MSSM increases monotonically with $\tanb$ \cite{2loop}, hence the
cross section here decreases correspondingly, for $\tanb \lsim 40$).

\begin{figure}[ht]
\bc
\epsfig{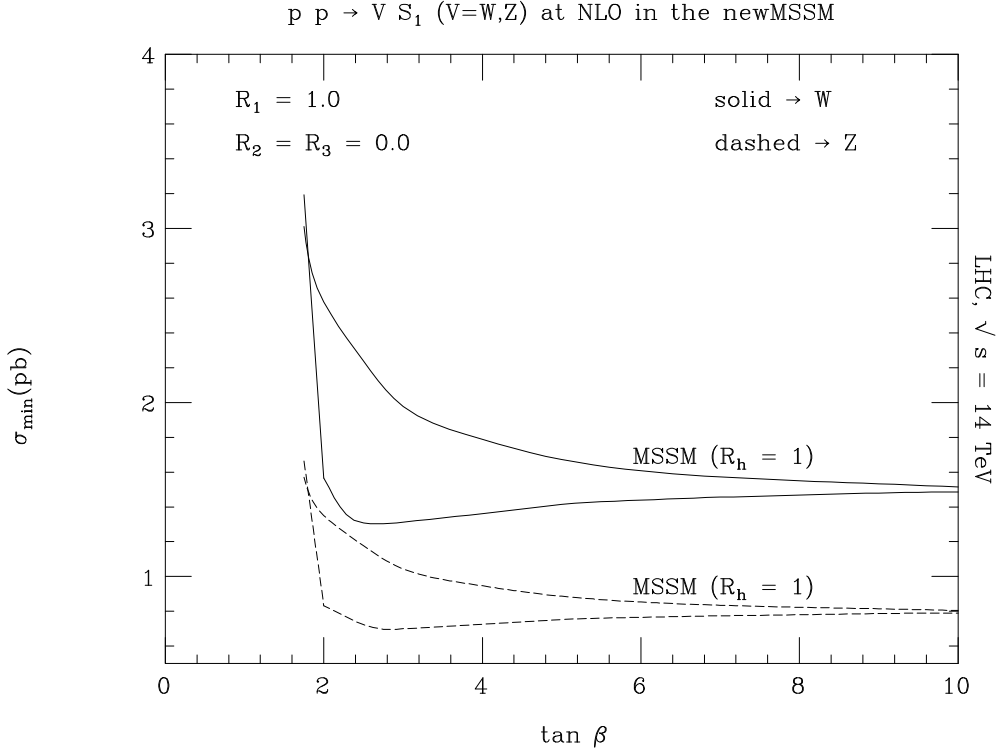}
\ec
\caption{Minimum of the production cross section of a neutral scalar Higgs
with SM-like couplings to $W^\pm$ and $Z$ gauge bosons at the LHC, as a
function of $\tanb$, in correspondence of the maximum values of its mass, in
both the MSSM and the nMSSM.}
\end{figure}

Thus, if $\tanb$ has already been measured in another context (e.g., in susy
sparticle processes), to detect an isolated scalar resonance in the $b\bar b$
channel (the dominant decay model in either model for
$M_{\mathrm{Higgs}}\lsim133.5$ GeV), at a rate well below the minimum one
predicted by the MSSM, could imply that physics beyond the latter would be
realized in nature. In fact, the reader should recall that the
BR$({\mathrm{Higgs}}\to b\bar b)$ is basically the same in both models
\cite{ioejames} and the combined uncertainties on the rates of the latter and
of the production modes (due to higher-order effects, parton distribution
functions, hard scale dependence, etc.) are of the order of just a few percent
\cite{wjszk}. In contrast, the differences between the MSSM and nMSSM rates
can be as large as a factor of two, in the vicinity of $\tanb \simeq 2.7$, and
well above the mentioned uncertainties for $\tanb$ up to 10 or so. Besides,
kinematical analysis of the $b\bar b$ system might provide further evidence in
this respect, if the mass resolution is larger than the difference between the
Higgs mass values, as predicted by the two models for a given $\tanb$. Similar
arguments can be made for the case of the $gg\to$~Higgs~$\to \gamma\gamma$
signature too, however, the production and decay phenomenology is here much
more involved (because loop processes take place at either stage), so that we
leave it aside for future consideration \cite{preparation}.

\section{Summary and conclusions}

We have studied the phenomenology of  the nMSSM, which promotes the
$\mu$-pa\-ra\-me\-ter into a singlet superfield, hence solving the well-known
$\mu$-problem of the MSSM. Besides, cubic self-couplings and possible
dimensionful couplings are avoided in the nMSSM thanks to a global discrete
R-symmetry, in turn broken by supergravity-induced tadpole corrections, which
solves both the so-called `domain wall' and `axion' problems. The new model is
truly minimal, in the sense that -- despite incorporating new fields -- it can
be parametrised by the same number of inputs as in the MSSM in the universal
case. Its phenomenology stands out quite different from that of both the
ordinary MSSM and the so-called Next-to-Minimal Supersymmetric Standard Model
(NMSSM) -- where the cubic self-interacting coupling is instead present. In
particular, the following aspects emerged as crucial from our analysis.

\bi
\item Assuming an accuracy up to the dominant top-stop contributions at
two-loop level and depending on the (reduced) couplings of the two lighter
CP-even Higgs bosons to the $Z$-boson, we have found that the upper limit on
the mass of the lightest Higgs state of the nMSSM can be 133.5 GeV, in
correspondence of the central value of the top mass (i.e., $m_t=173.8$ GeV),
that is, about 10 GeV higher than the MSSM value and within the Tevatron
Run-II reach. In addition, the upper limit on the mass of the lightest
'observable' Higgs boson (i.e., the next-to-lightest one, when the lightest
one couples invisibly to the $Z$-boson) could be as high as 160 GeV but still
within the LHC scope.

\item To remain with the Higgs sector, we also have described a benchmark
example that could allow one to phenomenologically distinguish the nMSSM
from the MSSM in the search for the lightest Higgs state at future hadron
colliders, such as the Tevatron Run-II and the LHC. If only one neutral Higgs
state is accessible through associated production with an EW gauge vector
$W^\pm$ or $Z$, via its $b\bar b$ decays, the knowledge of $\tanb$ and of the
production rate of such Higgs process could be enough to assign such a Higgs
state to one or the other of the two models, even prior to the investigation
of the mass resonance that can eventually be reconstructed from the $b\bar b$
system. Production and decay studies of all other Higgs states of the nMSSM
are now in progress~\cite{preparation}.

\item Despite the remarkable dissimilarities seen so far between the nMSSM
and the MSSM, one might quite rightly question that the phenomenology of the
nMSSM is very similar to that of the NMSSM, as far as the Higgs sector is
concerned. However, a dramatic difference is revealed between these two
models, if one investigates CP-violation effects. In fact, no matter the
actual value of $\lambda$, a peculiar feature of the nMSSM is the following:
Contrary to the NMSSM case, spontaneous CP-violation cannot occur in the Higgs
sector of the nMSSM, neither at tree-level nor at one-loop.

\item Finally, further differences between the nMSSM and any other model can
be appreciated in the sparticle sector. In fact, here, two concurrent aspects
render the phenomenology of the former both more `spectacular' and `natural',
in comparison to the MSSM and the NMSSM, respectively. Firstly, the lightest
neutralino appears to be an almost pure singlet state. Secondly, such a state
(the `singlino') is {\em always} the LSP of the theory, with a very small mass
(varying from a few MeV to a few GeV). The consequence is twofold. On the one
hand, in comparison to the NMSSM, the singlino LSP scenario requires no strong
constraints to be imposed on the parameter space. On the other hand, in
comparison to the MSSM, any sparticle decay chain involves a further step, the
NLSP $\to$ LSP transition, giving rise to additional cascades.

\item The stimulating issue that such an LSP could be a good dark matter
candidate or, alternatively, could be excluded from cosmological arguments,
also deserves attention. However, a quantitative analysis in this respect was
far beyond the intention of this note.
\ei

\noi Concluding, the nMSSM is, at the same time, the {\em best theoretically
motivated} and the {\em most economical} susy extension of the SM. Here, we have
pointed out differences or similarities between the new model, the NMSSM and
the MSSM, in both the Higgs and neutralino sectors. Future collider experiments,
at Tevatron Run-II and LHC, will be able to prove whether or not the nMSSM
is the realization of Supersymmetry that nature has chosen.

\bigskip\noi {\underbar{\bf Note}} While finalising this paper, whose main
results were made public already in Ref.~\cite{lisbon}, we became aware of
Ref.~\cite{Greeks}, by C. Panagiotakopoulos and A. Pilaftsis, which deals
with a similar subject. In this respect, although we agree with the more
general statements given in this other paper, we would like to remark that
there only the top-stop one-loop corrections were used, whereas here we have
calculated also the corresponding two-loop contributions. This
explains the significant discrepancy between their and our upper limit on the
lightest Higgs boson mass: $m_1<150$ GeV versus $m_1<133.5$ GeV.

\section*{Acknowledgements}
AD is supported from the Marie Curie Research Training Grant
ERB-FMBI-CT98-3438. CH and SM are grateful to the UK-PPARC for its financial
support. KT acknowledges travelling support from the TMR network " Beyond the
Standard Model". KT also thanks the `late' Theory Group at RAL for the kind
hospitality while part of this work was carried out.

\end{document}